\documentstyle[12pt]{article}
\begin{document}
\def\beqra{\begin{eqnarray}} \def\eeqra{\end{eqnarray}}
\def\beqast{\begin{eqnarray*}} \def\eeqast{\end{eqnarray*}}
\def\beq{\begin{equation}}	\def\eeq{\end{equation}}
\def\be{\begin{enumerate}}   \def\ee{\end{enumerate}}
\def\fnote#1#2{\begingroup\def\thefootnote{#1}\footnote{#2}\addtocounter
{footnote}{-1}\endgroup}
\def\ut#1#2{\hfill{UTTG-{#1}-{#2}}}
\def\sppt{Research supported in part by the
Robert A. Welch Foundation and NSF Grant PHY 9511632}
\def\utgp{\it Theory Group,  Department of Physics \\ University of Texas,
 Austin, Texas 78712}
\def\lag{\langle}
\def\rag{\rangle}
\def\rta{\rightarrow}
\def\haf{\frac{1}{2}}
\def\om{\omega}

\def\ijmp#1#2#3{{\it Int. J. Mod. Phys.} {\bf A#1,} #2 (19#3) } 
\def\mpla#1#2#3{{\it Mod.~Phys.~Lett.} {\bf A#1,} #2 (19#3) }   
\def\npb#1#2#3{{\it Nucl. Phys.} {\bf B#1,} #2 (19#3) }  
\def\np#1#2#3{{\it Nucl. Phys.} {\bf #1,} #2 (19#3) } 

\def\plb#1#2#3{{\it Phys. Lett.} {\bf B#1,} #2 (19#3) } 
\def\prc#1#2#3{{\it Phys. Rev.} {\bf C#1,} #2 (19#3) } 
\def\prd#1#2#3{{\it Phys. Rev.} {\bf D#1,} #2 (19#3) } 
\def\pr#1#2#3{{\it Phys. Rev.} {\bf #1,} #2 (19#3) } 
\def\prep#1#2#3{{\it Phys. Rep.} {\bf C#1,} #2 (19#3) } 
\def\prl#1#2#3{{\it Phys. Rev. Lett.} {\bf #1,} #2 (19#3) } 
\def\rmp#1#2#3{{\it Rev. Mod. Phys.} {\bf #1,} #2 (19#3) }

\parindent=2.5em

\ut{01}{96}


\begin{center}

\Large{\bf Collective Coordinates for D-branes }\normalsize 

\vspace{24pt}
Willy Fischler, Sonia Paban and Moshe Rozali\fnote{*}{\sppt} 

\vspace{12pt}
{\it \utgp}

\vspace{12pt}
\end{center}

\abstract{We develop a formalism for the scattering off D-branes that includes 
collective coordinates. This allows a systematic expansion in the string
coupling constant for such processes, 
including a worldsheet calculation for the
D-brane's mass.

\baselineskip=21pt
\setcounter{page}{1}
\input epsf.tex
\def\theequation{\thesection.\arabic{equation}}
\section{Introduction}

\indent\indent
In the recent developments of string theory, solitons play a central role. For 
example, in the case of duality among various string theories, solitons of a
weakly coupled string theory become elementary excitations of the 
dual theory
\cite{wit}. Another example, among many others, is the role of solitons in
resolving singularities in compactified geometries, as they become light
\cite{stro}.

In order to gain more insight in the various aspects of these recent 
developments, it is important to understand the dynamics of solitons in the
context of string theory. In this quest one of the tools at our disposal
 is the
use of scattering involving these solitons. This includes the scattering of
elementary string states off these solitons as well as the scattering among
solitons. An important class of solitons that have emerged are D-branes
\cite{dai}. In an influential paper \cite{pol1} it was shown 
that these D-branes
carry $R-R$ charges, and are therefore a central ingredient in the
aforementioned dualities. These objects are described by simple conformal field
theories, which makes them particularly suited for explicit calculations.

In this paper we use some of the preliminary ideas about collective coordinates 
developed in a previous paper \cite{fisc}, to describe 
the scattering of
elementary string states off  D-branes. We clarify and expand these ideas,
providing a complete analysis of this process. 

The paper begins by reviewing the presence of non-local violations of conformal 
invariance that are a consequence of the existence of translational zero modes.
It is then shown how the world sheet theory removes these anomalies through the
introduction of collective coordinates (which appear as ``wormhole
parameters''), and the inclusion of the contribution of disconnected 
worldsheets to the
S-matrix elements. As a byproduct of this analysis we calculate the mass 
of the soliton
by a purely worldsheet approach (that is amenable to higher order corrections).
We thereby recover a result obtained in \cite{dai} using effective low
energy considerations.

These ideas are then applied to the example of a closed string state scattering 
off a 0-brane. We show how to include the recoil of the brane, and 
discover the
interpretation of the wormhole parameters as collective coordinates. 

The conclusion contains some comments on higher order corrections to scattering 
and to mass renormalization as well as possible extensions of this work in
different directions. 

\section{Conformal anomalies.}
\setcounter{equation}{0}
\indent\indent
D-branes placed at different locations in space have the same energy. This 
implies the existence of spacetime zero modes, one for each spatial direction
transverse to the D-brane (i.e. one for each Dirichlet direction). 
Because of
these zero modes, the attempts to calculate scattering matrix elements in a weak
coupling expansion encounter spacetime infra-red divergences. In point-like
field theory these divergences arise from zero modes propagating 
in Feynman
diagrams (for example diagrams as in Fig. 1 , where $\eta_0$ is a zero mode).

\vspace{14pt}
\centerline{\epsfbox{wf1.fig}}

\vspace{14pt}

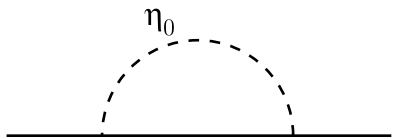
\centerline{Fig. 1: ~~Diagram containing the propagation of a zero-mode.}

\newpage

Such behavior can also be found in the canonical formalism, where it manifests 
itself by the presence of vanishing energy denominators, characteristic of
degenerate perturbation theory. This is resolved in quantum field 
theory by the
introduction of collective coordinates. The right basis of states, in which the
degeneracies are lifted, are the eigenstates of momenta conjugate to these
coordinates.

In string theory, however, we do not have a workable 
second quantized formalism. Instead, we have to rely on a first 
quantized description. The weak coupling expansion consists of 
histories spanning worldsheets of successively higher topologies. Yet, the
aforementioned divergences are still present, and are found in a specific corner
of moduli space. The divergences appear as violations of conformal invariance, as
described below. To the lowest order, these singularities are found when one
consideres worldsheets
as depicted in Fig. 2.

\vspace{14pt}
\centerline{\epsfbox{wfig2.fig}}

\vspace{12pt}
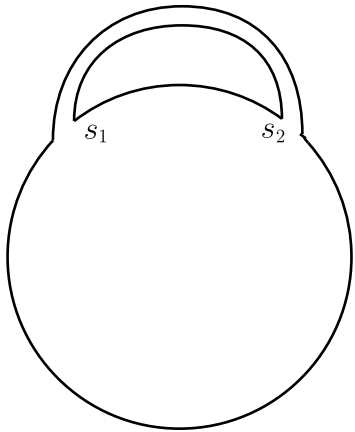
\centerline{Fig. 2:~~ The surface $\sum$.} 

\vspace{12pt}
\noindent This surface,
$\sum$, is obtained by attaching a very long and thin strip to the boundary of 
the disc. This is equivalent to 
removing two small segments of the boundary of
the disc and identifying them, thus creating a surface with the topology of an
annulus. In the degeneration limit, when the strip becomes infinitely long and
thin, the only contribution comes from massless open 
string states propagating
along the strip. In the case of Dirichlet boundary conditions, the only such
states are the translational zero modes. 

In the case of closed string solitons, the divergence appears for degenerate 
surfaces containing long thin handles, along which the zero mode, a closed
string state in that case, propagates. In both cases these limiting
world-histories are reminiscent of the field theory Feynman diagrams, as in 
Fig. 1 . 

We are now ready to properly extract the divergences. As discussed in 
\cite{pol2} for the closed bosonic string theory, 
the matrix element in the
degeneration limit can be obtained by a conformal field theory calculation on
the lower genus surface, with operators inserted on the degenerating segments
(see Fig. 3 ). 

\vspace{18pt}
\centerline{\epsfbox{wffig4.fig}}

\vspace{18pt}
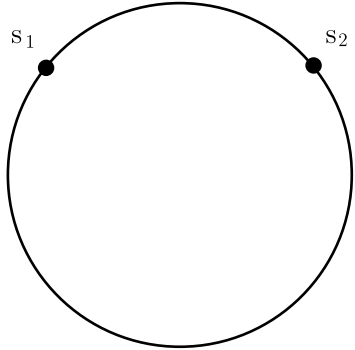
\centerline{Fig. 3:~~ The degenerations limit of the surface $\sum$.} 

\vspace{14pt}

This can be expressed formally
as follows:
\beq
\lag V_1\ldots V_n\rag_\Sigma =\sum_a\int\frac{dq}{q}\;q^{-2h_a}\lag 
\phi_a(s_1)\,\phi_a(s_2)\;V_1\ldots V_n\rag_{{\rm Disc}} 
\eeq 
where $\{\phi_a\}$
is a complete set of eigenstates of $L_0$ (which generates translations along
the strip), with weights $h_a$. The limit $q\rta 0$, therefore, 
projects the sum
into the lowest weight states.

As an illustration, it is useful to first consider the same limit for the usual 
open bosonic string (with Neumann boundary conditions). 
The lowest weight states
(excluding the tachyon) are:
\beq
\phi_k^i=N_k \sqrt{g}\;\left(\frac{ dX^i}{ds}\right)\; e^{ikX} 
\eeq 
with 
$h_k=\frac{k^2}{2}$ and $N_k$ a normalization factor. 
Using the above formula (2.1), one
gets: 
\beq
\lag V_1\ldots V_n\rag_\Sigma =\sum_i \int \frac{d^dk}{(2\pi)^d}\int dq\,
q^{-(1+k^2)} \, \left\lag\phi_k^i(s_1)\phi_k^i(s_2)\,V_1\ldots
V_n\right\rag_{{\rm Disc}}\,. 
\eeq 

This expression exhibits the exchange of a photon, where $\log q$ is the time 
parametrizing the photon worldline. The requirement of 
factorization fixes the
normalization of the operators
$\phi_a$.

We now turn to the analysis for the D-branes, which is rather similar. The 
Dirichlet p-brane is defined by imposing Neumann boundary conditions on
$X^N=X^0\ldots X^p$, and Dirichlet boundary conditions 
on the remaining
coordinates,
$X^D$. The lightest states (excluding the tachyon) 
are the translational zero modes of
the  D-brane:
\beq
\phi_i = N\,\partial_n  X^D_i\,e^{i\om X_0}\qquad i=p+1,\ldots, d-1\;. 
\eeq
where $X_0$ is target time, $\partial_n$ is the derivative normal to the 
boundary, and $N$ is a normalization constant. The existence 
of these zero modes 
reflects the degeneracy of different D-branes configurations, obtained by rigid
translations $(\vec{k}_N=0)$ in the Dirichlet directions. 

In order to allow a general p-brane $(p>0)$ to recoil one needs to compactify 
the spatial Neumann directions and wrap the p-brane around them. In that case
the mass of the p-brane is finite and proportional to the spatial 
Neumann
volume. For the sake of clarity we focus on the scattering off a 0-brane, where
no such compactification is necessary. 

The normalization constant $N$ that appears in (2.4) is determined by 
factorization; consider the S-matrix element involving two closed string
tachyons whose vertex operators are:
\beqra
V_1 &=& N_t\int d^2z_1\,e^{i\om_1 X_0(z_1)-i\vec k_1^D\cdot\vec X^D(z_1)} 
\nonumber \\ V_2 &=& N_t\int d^2z_2\,e^{i\om_2 X_0(z_2)-i\vec k_2^D\cdot\vec
X^D(z_2)}\,. \eeqra

The factorization on open string poles is obtained by letting one of the 
tachyons approach the boundary (see Fig. 4).

\vspace{18pt}
\centerline{\epsfbox{newfigs.fig}}

\vspace{18pt}
\centerline{Fig. 4:~~ Factorization on the Disc.} 

\vspace{14pt}

In the factorization one recognizes the various open string poles, in 
particular the tachyon at $\om^2=-\haf$, and the 
zero mode at $\om^2=0$. This yields
the normalization factor for the zero 
mode:\footnote{In our notation the open string
tachyon is
$$\haf\,\sqrt{g}\,e^{i\om X_0}$$} 
\beq
 N=\frac{\sqrt{g}}{4}\;.
\eeq

Since these states don't carry Dirichlet momenta, they are discrete quantum 
mechanical states.

We are now ready to extract the violations of conformal invariance for S-matrix 
elements calculated on the annulus. As mentioned above, 
these divergences appear
in the limit of moduli space described in Fig. 2 . In this limit the divergent
contribution to an S-matrix element involving $n$ elementary string states is:
\beq 
\left\lag \int V_1\ldots\int V_n\right\rag_\Sigma=\frac{-g}{16T}\,
\log\,\epsilon
\left\lag\partial_n\vec{X}_D(s_1)\cdot \partial_n \vec X_D(s_2)\int V_1\ldots
 \int V_n\right\rag_{\rm Disc} 
\eeq
where $T$ is a large target time cut-off, and $\epsilon$ is a 
world-sheet cut-off.
The  appearance of $T$ is necessary in order to properly extract the divergent
contributions due to the translational zero modes.

\section{Wormhole Parameters}
\setcounter{equation}{0}
\indent\indent In order to restore conformal invariance we appeal to the 
Fischler-Susskind mechanism \cite{fisc2}. The idea is to modify the
two-dimensional action such that the sum of the various world-sheet
contributions to the S-matrix element is conformally invariant.

The Fischler-Susskind mechanism requires in this case the addition 
of a non-local 
operator $O$ to the world-sheet action:
\beq
O=\frac{g}{8T}\,\log^2 \epsilon \int ds_1\,\partial_n\vec X_D(s_1)\int d
s_2\,\partial_n\vec X_D(s_2)\,.
\eeq
This operator is a bilocal involving the translational zero mode vertex 
operators.   Notice the additional logarithmic divergence beyond the one that is
already present in (2.7). This additional divergence 
accounts for the volume of
the residual M\"obius subgroup that fixes the locations of two vertex operators:
\vfill
\pagebreak
\beqra
&&\frac{1}{V_{{\rm M\ddot{o}bius}}}\, 
\left\lag\int ds_1\partial_n \,X(s_1)\int ds_2 \,X
(s_2)\int V\ldots\int V\right\rag_{{\rm Disc}} = \nonumber \\ 
&&\frac{1}{V_{{\rm Residual}}} \,\left\lag 
\partial_n X (s_1)\,\partial_n\,X(s_2) \int
V\ldots\int V\right\rag_{{\rm Disc}} 
\eeqra
where $V_{{\rm Residual}}=-2\log \,\epsilon$~ 
is the residual M\"obius volume as defined above \cite{jl}.
In the case of the Fischler-Susskind mechanism involving a local 
operator this would
correspond instead to a finite factor. 

In order to rewrite the sum of histories as an integral weighted by a local 
action we introduce wormhole parameters $\alpha_i$, one for each Dirichlet
direction. As will become clear these parameters 
are proportional to the momenta
conjugate to the center of mass position of the 0-brane.

The S-matrix element then takes the form: 
\beqra
S &=& \int d\vec \alpha e^{-\vec\alpha^{2}/2}\,\int DX\;e^{I}\int 
V_1\ldots \int V_n \\[6pt] 
I &=& \frac{-1}{4\pi}\,\int d^2 z (\partial\vec
X\cdot\vec\partial\vec X+\partial X^0\cdot\partial X^0)+
\frac{\vec\alpha\sqrt{g}}{2\sqrt{T}}\, 
\log\,\epsilon\int ds\; \partial_n \vec X
\nonumber \\ 
&&\qquad +\; \frac{1}{8\pi}\,\int ds \;\vec a(X_0)\partial_n\vec X
\eeqra 
As will be shown below, the inclusion of the last term in the action is
important in restoring conformal invariance. At the lowest order in 
the string
coupling constant $g$: 
$$ \vec a(X_0)=\vec a_0 $$ 
where $\vec a_0$ is a constant
(independent of $X_0$), and represents the fixed location of the brane.

\section{Scattering off a 0-brane}
\setcounter{equation}{0}
\indent\indent 
In organizing various contributions that violate conformal 
invariance we begin by considering the ``vacuum persistence amplitude": 
\beqra
_{\phantom{\hat{\dot T}-}T/2}\lag 0\,|\,0\rag_{-T/2}^{\vec\alpha,
\vec{a}} = \int DX\,\exp\Bigg\{
\frac{-1}{4\pi}\,\int \partial X \bar\partial X &+&
\frac{\vec\alpha
\sqrt{g}}{2\sqrt{T}}\,\log\,\epsilon \int\partial_n\vec X +
\nonumber \\ 
&+& \frac{1}{8\pi} \int \vec a(X_0)\partial_n\vec X \Bigg\}\,.
\eeqra

This expression differs from the vacuum persistence amplitude in the absence of 
$\vec\alpha,\,\vec{a}$, as can be seen in the following expression:
\beqra
_{\phantom{\hat{\dot T}-}T/2}\lag 0\,|\,0\rag_{-T/2}^{\vec\alpha,\vec{a}}& 
= & \exp\left(-
\frac{\alpha^2}{T}\,\log\,\epsilon\right)\,\exp\left(
\frac{\dot{a}^2}{\pi^2g}\,\log\,
\epsilon\right)
\nonumber \\[5pt] 
&& \exp \left(-\frac{1}{2\pi\sqrt{g T}}\, \vec a_0\cdot\vec \alpha\right)
{\phantom{\rag}_{\phantom  {\hat{\dot T-}}T/2}\left\lag
0\,|\,0\right\rag_{-T/2}^{\vec\alpha=0,\vec a=\vec{a}_0}} 
\eeqra  
In order to  explain equation (4.2), we note first that this contribution to the
S-matrix element shows up diagrammatically as the 
sum of  disconnected diagrams   (see
Fig. 5). These disconnected diagrams have to be added 
to the more conventional connected
diagrams in order to restore conformal 
invariance\footnote{Disconnected diagrams
have previously been considered in the case of the D-instanton \cite{pol3}.}

\vspace{14pt}
\centerline{\epsfbox{newfig2.fig}}

\vspace{14pt}

\centerline{Fig. 5: ~~ The sum over disconnected diagrams.} 

The various terms appearing in the exponent in eq. (4.2) are calculated on the 
disk as the two-point function of the operator $O'$: 
$$O'= \int
ds\,\left(\frac{\vec\alpha\sqrt{g}}{2\sqrt{T}}\, \log\,\epsilon\,
\partial_n\vec
X + \frac{1}{8\pi}\, \vec a(X_0)\,\partial_n \vec X\right) \;.$$ 
Care has to be
exercised to carefully account for the M\"obius volume associated to 
the disc.
The S-matrix element then becomes: 
\beqra 
S&=&\int d\vec\alpha\,e^{-\vec\alpha^2/2}\int
DX\,\exp\left(-\frac{\vec\alpha^2}{T}\,\log
\epsilon\right)\,    \exp\left( \frac{\vec{\dot{a}}^2}{\pi^2g} \,\log
\epsilon\right) \nonumber \\ 
&& \exp\left(-\frac{1}{2\pi\sqrt{gT}}\,\vec
a_0\cdot\vec\alpha\right)\, e^{I} \int V_1\ldots \int V_n\,.
\eeqra

For simplicity we analyze the scattering of a closed string tachyon off the 
D-brane. The initial and final momenta are respectively $\vec k_1$ and $\vec
k_2$ with excess momentum $\vec P=\vec k_1 +\vec k_2$.

The S-matrix element (4.3) is then, after some straightforward algebra:  
\vfill
\pagebreak

\beqra
S &=& \int d\vec\alpha\,e^{-\vec\alpha^2/2}\,\exp\left\{
\frac{\vec a_0}{2}\cdot \left(i
\vec P-\frac{\vec\alpha}{\pi\sqrt{gT}}\,\right) \right\} \nonumber \\ 
&&\exp\left\{ -\left(\frac{\vec\alpha}{\sqrt{T}} -i\pi\sqrt{g} 
\vec P\right)^2\,\log\,
\epsilon\right\} \nonumber \\
&& \exp\left\{  \left( \frac{\vec{\dot{a}}^2}{\pi^2 g}\,- \pi^2g\,\vec P^2
\right) \,\log\,\epsilon\right\} \nonumber \\ 
&& \exp\left\{ \haf\left(
\om_1+\om_2+\haf \,\vec{\dot{a}}\cdot\vec P\right)^2\,
\log\,\epsilon\right\}\;F(\vec\alpha,\vec k_1,\vec k_2,\vec{\dot a}) 
\eeqra
where $F$ is the finite part of the S-matrix element, up to order
$g^0$, evaluated using the action (3.4).

 Conformal invariance is restored only if: 
\beqra
&&\vec\alpha = i\pi\vec P\,\sqrt{gT} \\
&&\vec{\dot a} = \pi^2 g\vec P \\
&&\om_1+\om_2 +\haf\, \vec{\dot{a}}\cdot\vec P=0
\eeqra
equation (4.5) identifies $\vec{\alpha}$ as being 
proportional to the recoil 
momentum; equation (4.6) identifies $\vec{\dot a}$ as the recoil velocity
(interpreting the coefficient as the inverse mass); equation (4.7) expresses the
conservation of energy; using (4.6) it can be written as:
\beq
\om_1+\om_2 +\frac{\vec P^2}{2M}=0\,. 
\eeq
This can also be obtained from (4.4) by 
integrating over the constant part of $X_0$.
The S-matrix element then becomes:
\beq
 S=\exp\left(-\haf\,\pi^2g\,\vec P^2T\right)\,
\delta(\om_1+\om_2 +\haf\,\pi^2
g \vec P^2)\,G(\vec k_1,\vec k_2) 
\eeq
where $G(\vec k_1,\vec k_2)$ is obtained from $F$ 
above by substituting the above
values for $\vec\alpha,\vec{\dot{a}}$, and contains in particular the 
contribution from intermediate, virtually recoiling, brane. 

Note that relation (4.6), that enforces momentum conservation, eliminates the 
dependence on $\vec{a}_0$.\footnote{Similarly, the conformal 
violations proportional to
$\vec{\ddot a}$ cancel, at this order,  when adding 
connected and disconnected diagrams.}
We therefore conclude that (4.10) represents the transition 
from initial D-brane at rest,
to a final state in which the D-brane has recoil momentum $\vec P$. 
This allows us to identify: 
$$M=\frac{1}{\pi^2g}$$
which is consistent with the results of \cite{dai}.

\section{Conclusion}
\setcounter{equation}{0}
\indent\indent  
In conclusion, we have applied the Fischler-Susskind mechanism to the
case of  D-brane recoil. This approach can be systematically extended 
to higher orders in
string perturbation theory. Contributions from disconnected diagrams have to be
included in order to restore conformal invariance and translational invariance.
A purely two-dimensional calculation of the soliton mass is obtained.  

In this paper we consider scattering of elementary string state 
off the D-brane.  This
still leaves the question
 is the scattering among D-branes \cite{cb}. Since D-branes are BPS
saturated one might hope to find the metric on the moduli space 
describing such
scattering ( in analogy to the work of \cite{ati} on BPS monopoles). 
This would entail
finding the space on which the wormholes move. 

Another interesting question is
the treatment of fermionic collective coordinates, from a two-dimensional
viewpoint, and the appearance of a manifestly supersymmetric collective
coordinate description. 

In summary, a better understanding of the dynamics of
D-branes can shed more light on the relations among string theories. 

\section*{Acknowledgments}

\indent\indent We thank D. Berenstein, S. Chaudhuri, M. Dine, and J. Distler 
for discussions. This work is supported in part by the Robert A. Welch 
Foundation, and NSF grant PHY9511632.  

After the completion of this work
we have received a paper dealing with similar issues \cite{vp}. 
We disagree with the
conclusions of that paper.

\end{document}